\documentclass{article}
\usepackage[preprint]{spconfa4}
\usepackage{amsmath,graphicx,amsfonts}

\usepackage{blindtext,lipsum,letltxmacro,xparse}
\usepackage[table]{xcolor}
\usepackage{comment}
\usepackage{booktabs}
\usepackage{subcaption}
\usepackage{tikz}
\usepackage{tikzsymbols}
\usetikzlibrary{positioning}
\usetikzlibrary{shapes}
\usetikzlibrary{calc}

\title{Analysis of impact of emotions on target speech extraction\\and speech separation}

\name{\textit{J\'{a}n \v{S}vec$^1$},
\textit{Kate\v{r}ina \v{Z}mol\'{i}kov\'{a}$^1$},
\textit{Martin Kocour$^1$},
\textit{Marc Delcroix$^2$},
\textit{Tsubasa Ochiai$^2$},\\
\textit{Ladislav Mo\v{s}ner$^1$},
\textit{Jan ``Honza'' \v{C}ernock\'{y}$^1$}}

\address{$^1$Brno University of Technology, Faculty of IT, IT4I Centre of Excellence \\ 
 $^2$NTT Corporation, Japan}
 
\begin{document}
\ninept

\maketitle

\begin{abstract}
Recently, the performance of blind speech separation (BSS) and target speech extraction (TSE) has greatly progressed. Most works, however, focus on relatively well-controlled conditions using, e.g., read speech. The performance may degrade in more realistic situations. One of the factors causing such degradation may be intrinsic speaker variability, such as emotions, occurring commonly in realistic speech. In this paper, we investigate the influence of emotions on TSE and BSS. We create a new test dataset of emotional mixtures for the evaluation of TSE and BSS. This dataset combines LibriSpeech and Ryerson Audio-Visual Database of Emotional Speech and Song (RAVDESS). Through controlled experiments, we can analyze the impact of different emotions on the performance of BSS and TSE. We observe that BSS is relatively robust to emotions, while TSE, which requires identifying and extracting the speech of a target speaker, is much more sensitive to emotions. On comparative speaker verification experiments we show that identifying the target speaker may be particularly challenging when dealing with emotional speech. Using our findings, we outline potential future directions that could improve the robustness of BSS and TSE systems toward emotional speech.
\end{abstract}


\section{Introduction}
\label{sec:intro}
Speech processing applications often suffer from reduced performance in real-world environments because of the presence of interfering speakers. Recent research tackles this problem by pre-processing the multi-talker signal to isolate the speech of individual speakers using deep learning approaches. There are two dominant approaches: blind speech separation (BSS)~\cite{wang2018supervised,hershey2016deep,yu2016permutation,luo2019convtasnet,subakan2021sepformer} and target speech extraction (TSE)~\cite{zmolikova2019speakerbeam,wang2020voicefilter}. The task of BSS is to estimate all sources in a mixture; this is needed in some applications such as automatic meeting transcription~\cite{araki2007blind,raj2021integration}. In contrast, TSE aims to extract the speech signal of a target speaker only, while removing all other interferences. It can be a practical alternative to BSS for applications such as smart speakers~\cite{wang2020voicefilter}. The target speaker is determined by providing, e.g., an enrollment recording of the voice of the target speaker. In contrast with BSS, TSE needs to perform both separation and identification of the target speaker.

BSS and TSE have significantly progressed with the advent of deep learning and can achieve excellent separation and extraction performance in well-controlled conditions. However, the performance sometimes degrades when tackling more realistic conditions. Understanding the cause of the performance degradation is crucial to further progress research on BSS and TSE.

There are different factors that may influence the performance of TSE and BSS. Noise and reverberation have been shown to affect the separation performance significantly~\cite{wichern2019wham,maciejewski2020whamr}. Some works also explored the effect of voice characteristics and showed the difficulty of performing TSE or BSS on mixtures of speakers with similar voices~\cite{ditter2019influence,delcroix2020improving}. The language~\cite{borsdorf2021globalphone} or environment~\cite{maciejewski2019analysis} mismatch has also been shown to be detrimental to the BSS performance. These prior works have focused on the impact of external conditions (noise or reverberation) or global speaker characteristics (such as voice characteristics and
languages). However, the voice of a speaker changes also due to, e.g., health condition, type of speech (read, presentation, natural conversations), emotions etc. The impact of such intrinsic speaker variability has been less explored. In this paper, we focus on the impact of emotion on BSS and TSE. Understanding the effect of emotions on BSS and TSE tasks can have crucial implication on the design of applications as we would expect that voice-user interfaces to understand us even when we, for example, become angry at them.

Several studies have shown that the~emotional state of the~speaker influences the~articulation~\cite{lee05c_interspeech} and the~prosody~\cite{scherer2003vocal}. These changes in the~speech signal lead to decreased performance of various speech technologies, including speech recognition~\cite{rushab2019asr_emo} or speaker verification~\cite{pappagari2020xvectors}. These findings suggest that emotions might influence also the~performance of TSE and BSS. There are several ways how the emotions can affect these systems. First, the separation or extraction performance may degrade because BSS or TSE models may not represent well the characteristics of emotional speech if emotional data was not well represented in the training data. On the other hand, it may be easier to separate or distinguish speakers in the mixture if they speak with different emotions. Finally, for TSE, the target speech and the enrollment could have different emotions, which may make it hard to identify the target speaker in the mixture.

In our work, we analyze the effect of emotions on TSE and BSS systems. For achieving this, we create a dataset of emotional speech mixtures,  based on The Ryerson Audio-Visual Database of Emotional Speech and Song (RAVDESS)~\cite{livingstone2018ravdess}. We call this dataset RAVDESS2Mix. The dataset is prepared in a way, that enables us to isolate the impact of emotions from other factors such as semantic content. With this dataset, we reveal that BSS is not affected much by emotions. In contrast, TSE performance degrades severely when the enrollment and target utterance in the mixture have mismatched emotions. We thus hypothesize that the challenging task is the identification of the target speaker using mismatched enrollment, and we support this hypothesis with speaker identification experiments.

\begin{figure*}[t]
    \centering
    \input{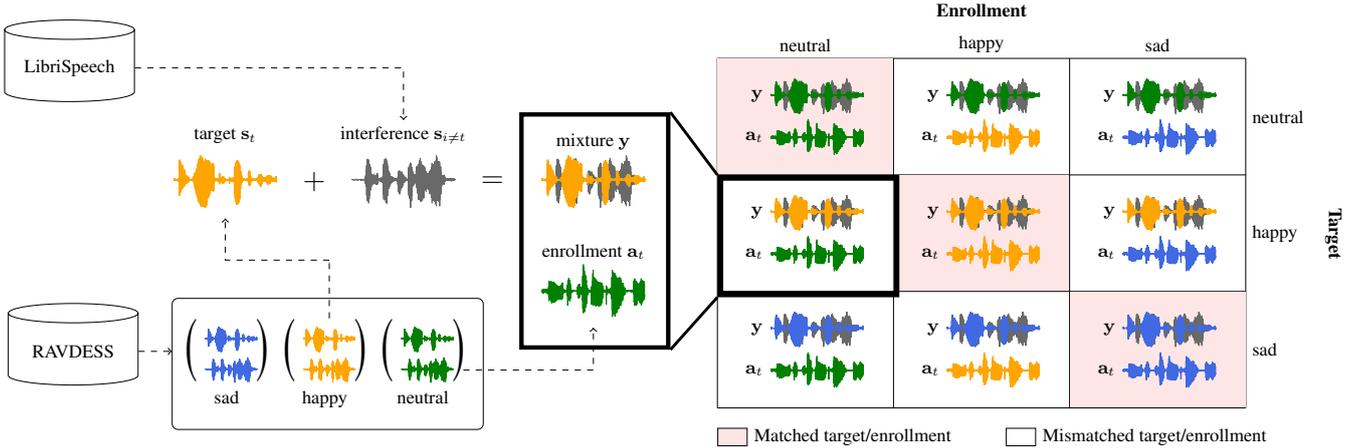}
    \caption{RAVDESS2Mix: illustration of the dataset creation for 3 emotions, showing the detailed generation of one sample of neutral-happy set. In actual RAVDESS2Mix, we use 8 different emotions, leading to $8\times 8 =64$ combinations of $(\text{mix},\text{enroll})$.}
    \label{fig:dataset}
\end{figure*}

\section{BSS and TSE tasks description}
\label{sec:task}

In both TSE and BSS tasks, the~input of the~processing is a~mixture defined as
\begin{equation}
    \label{eq:mixture}
    \mathbf{y} = \sum_{i=1}^{S} \mathbf{s}_i + \mathbf{n},
\end{equation}
where $\mathbf{y} \in \mathbb{R}^T$, $T$ is the~duration of the signals in samples, $\mathbf{s}_i$ is the time-domain speech signal of the $i$-th speaker in the~mixture, $S$ is the~number of speakers and $\mathbf{n}$ represents additional noise. 

A BSS model takes the~mixture $\mathbf{y}$ as input and aims to obtain the~signals of all $S$ speakers as the~output,
\begin{equation}
    \label{eq:mask_sep}
    \{\hat{\mathbf{s}}_1,\dots,\hat{\mathbf{s}}_S\} = f^{(sep)}(\mathbf{y}), \\
\end{equation}
where $\hat{\mathbf{s}}_i$ is the~estimated signal of the $i$-th speaker. In this work, we implement the~BSS model $f^{(sep)}$ with a fully-convolutional time-domain audio separation network (Conv-TasNet) \cite{luo2019convtasnet}, that has been widely used for speech separation.

TSE also accepts the~mixture $\mathbf{y}$ as the~input but, in contrast with BSS, also expects additional information in the~form of an enrollment utterance of the~target speaker. We denote this utterance for target speaker $t$ as $\mathbf{a}_t \in \mathbb{R}^{T'}$ where $T'$ might be different from $T$. TSE aims to estimate the~speech of speaker $t$ at the~output. This is done in two steps: first, a speaker embedding $\mathbf{e}_t$ is extracted from the enrollment utterance, and second, the embedding is used to inform the extraction process about the target speaker:
\begin{align}
  \label{eq:mask_extraction}
  \mathbf{e}_t &= f^{(emb)}(\mathbf{a}_t) \\
  \hat{\mathbf{s}}_t &= f^{(tse)}(\mathbf{y}, \mathbf{e}_t),
\end{align}
where $f^{(emb)}$ and $f^{(tse)}$ represent the speaker embedding computation and speech extraction network, respectively. The speaker embedding is then used to inform the extraction network, $f^{(tse)}$, about the~target speaker.
The~TSE formulation of the~problem avoids some issues of speech separation. For instance, the~number of speakers in the~mixture does not need to be known. Also, the~speaker information clearly determines the~speaker at the~output, therefore the~global speaker ambiguity is avoided~\cite{zmolikova2019speakerbeam}. However, the~TSE processing $f^{(tse)}$ needs to perform the~identification of the~target speaker, in addition to the~separation itself, which can lead to additional errors.

We follow time-domain SpeakerBeam \cite{delcroix2020improving} to implement $f^{(emb)}$ and $f^{(tse)}$. Time-domain SpeakerBeam is based on the~Conv-Tasnet architecture, extending it to accept the speaker embedding. This is done by performing an element-wise multiplication of the~embedding $\mathbf{e}_t$ with the series of input signal hidden representations in the~extraction network, $f^{(tse)}$. With SpeakerBeam, $f^{(emb)}$ is implemented as a neural network, which is jointly trained with the extraction network. 

\section{Proposed framework to analyze the impact of emotions}
\label{sec:dataset}

\subsection{Controlled dataset design}
\label{sec:controlled_dataset_design}

It may be hard to disentangle the effect of emotions from other factors that may affect separation or extraction performance such as the interference signal, the semantic content of the target or enrollment utterance using existing corpora. 
Here, we design a controlled experimental framework to analyze specifically the effect of the emotional speech on TSE and BSS tasks. 
This is done by creating mixtures of an emotional target speaker and a relatively neutral interfering speaker. We create mixtures with identical parameters (speaker identities, interference signal, mixing SNR, semantic content etc), and vary only the emotions of the target speaker. This will enable to compare BSS and TSE performance across emotions.

For TSE, we also use enrollment with various emotions. This allows us to analyze the effect of the emotional variability, both in cases when the enrollment and target have \textit{matched}  and \textit{mismatched} emotions.

We created only emotional mixtures for evaluation and intentionally used Librimix~\cite{cosentino2020librimix} for training. 


\subsection{Dataset creation}
\label{sec:dataset_creation}

We create the~RAVDESS2Mix dataset of two-speaker mixtures, such that it is possible to isolate the~effect of the~emotion on the~BSS and TSE tasks. We designed this dataset so that it relies on openly available resources to allow reproducibility~\footnote{Data generating code is available at https://github.com/BUTSpeechFIT/RAVDESS2Mix.}.

For the emotional utterance in the dataset, we use RAVDESS~\cite{livingstone2018ravdess}. The~database contains 24~professional actors (12~female, 12~male), expressing acoustically and visually two semantic statements. Utterances are spoken with eight different emotions (neutral, calm, happy, sad, angry, fearful, surprise, disgust) with a~North American accent. Each emotion is produced at two levels of emotional intensity (normal, strong). Thus, by utilizing RAVDESS, we can create mixtures that differ only in the emotions of the target speaker. The acoustic part of the~corpus contains also songs, but we use only speech. For the interferences, we use utterances taken from the test set of LibriSpeech~\cite{panayotov2015librispeech}\footnote{Note that we use LibriMix for training, which prevents the systems from learning to discriminate the target and interference simply because the emotional speech and the interference are taken from different datasets.}. 

Figure~\ref{fig:dataset} shows the mixture creation process of RAVDESS2Mix. We created 384 mixtures for each emotion (each cell in the grid in Figure~\ref{fig:dataset}). The RAVDESS dataset has 24 speakers, two semantic statements per speaker, two repetitions per statement and two levels of emotional intensity which results in $24 \times 2 \times 2 \times 2 = 192$ unique utterances. Mixing these with different interfering utterances from LibriSpeech resulted into 384 mixtures.

For TSE, we use as enrollment the utterances with different semantic statement than that in the mixture. The enrollment utterance also has one of eight different emotions. There are thus eight variants of the mixtures that differ only by the emotion of the enrollment. This creates 64 ($8 \times 8$) variants of the pair $(\text{mix}, \text{enroll})$.

\section{Experiments}
\label{sec:experiments}

\subsection{Configuration of TSE and BSS}
\label{sec:configuration_tse_bss}



The TSE and BSS models were trained using the~Asteroid~\cite{pariente2020asteroid} toolkit. For BSS, we used Conv-TasNet with the following hyper-parameters: 8 blocks, 4 repetitions, and block dimensionality (128, 512, 128), where the output is produced by skip connections. For TSE, we used the implementation of SpeakerBeam provided in our GitHub repository\footnote{https://github.com/BUTSpeechFIT/speakerbeam} which is based on Conv-TasNet architecture. We use the same hyper-parameters as in BSS. The auxiliary network consists of eight connected 1-D Convolutional blocks~\cite{luo2019convtasnet}, corresponding to one repetition in Conv-Tasnet architecture. The auxiliary network has its own encoder. The speaker information is used after the first repetition of the extraction network. For training, we dynamically created mixtures from LibriSpeech utterances with noise from WHAM~\cite{wichern2019wham}, closely following the mixing process used in LibriMix dataset \cite{cosentino2020librimix}. For both TSE and BSS, we set the batch size to 20, number of epochs to 200, and used the Adam optimizer~\cite{kingma2014adam} with an initial learning rate of 1e-3, which is halved after five consecutive epochs with no reduction in validation loss. We worked with 8~kHz sample rate to reduce computational cost.



\begin{table}[t]
    \centering 
    \caption{Overall performance of the BSS and TSE systems in terms of SDR improvements [dB] over the mixture SDR (first row).}
    \label{tab:librimix_orig}
    \begin{tabular}{cc c c c}\toprule
      & \multicolumn{1}{c}{Libri2Mix} & \multicolumn{3}{c}{RAVDESS2Mix} \\\cmidrule{3-5}
      && Avg. over & target/enroll & target/enroll\\
      &&  all sources & matched & mismatched\\\midrule
      mix & -0.2 & 0.2 & \multicolumn{2}{c}{-3.2} \\\midrule
      BSS & 15.1 & 15.5 & \multicolumn{2}{c}{14.4} \\
      TSE & 14.8 & - & 14.1 & 12.4 \\
      \bottomrule
    \end{tabular}
\end{table}
\subsection{Results}
\begin{table*}[t]  
  \caption{SDRi [dB] for the estimated target source with BSS and TSE systems. 
  }
  \label{tab:emo_bss_tse}
  \centering
  \label{tab:normal_intensity}
  \begin{tabular}{l c c c c c c c c c }
    \toprule
    & BSS & \multicolumn{8}{c}{TSE (Enrollment emotion)} \\\cmidrule{3-10}
    Target emotion &  & \textit{neutral} & \textit{calm}  & \textit{happy} & \textit{sad}   & \textit{angry} & \textit{fearful} & \textit{disgust} & \textit{surprised} \\\midrule
    \textit{neutral}   & 15.0 & \cellcolor{red!10}15.1 & 14.8 & 12.0 & 13.3 & 9.5 & 9.8 & 13.2 & 12.8 \\
    \textit{calm}      & 14.5 & 13.6 & \cellcolor{red!10}14.1 & 9.2 & 10.8 & 6.8 & 6.7 & 10.9 & 10.6 \\
    \textit{happy}     & 14.5 & 13.0 & 12.5 & \cellcolor{red!10}14.5 & 13.4 & 13.8 & 13.1 & 13.8 & 13.6 \\
    \textit{sad}       & 14.3 & 13.2 & 12.7 & 12.7 & \cellcolor{red!10}13.7 & 11.3 & 11.9 & 13.0 & 12.8 \\
    \textit{angry}     & 14.6 & 12.3 & 11.3 & 14.3 & 12.7 & \cellcolor{red!10}14.5 & 13.8 & 13.8 & 13.4 \\
    \textit{fearful}   & 14.5  & 12.1 & 12.0 & 13.9 & 13.3 & 13.6 & \cellcolor{red!10}14.2 & 13.4 & 13.8 \\
    \textit{disgust}   & 14.0 & 12.4 & 12.0 & 13.0 & 12.5 & 12.0 & 11.8 & \cellcolor{red!10}13.6 & 12.7 \\
    \textit{surprised} & 13.9 & 13.0 & 12.7 & 13.4 & 12.3 & 12.5 & 12.6 & 13.1 & \cellcolor{red!10}13.7 \\
    \bottomrule
  \end{tabular}
\end{table*}

First, we compare both TSE and BSS results on RAVDESS2Mix with those on the test-set of Libri2mix without noise \cite{cosentino2020librimix}, to get an initial idea of how challenging the emotional dataset is. Table~\ref{tab:librimix_orig} shows the signal to distortion ratio improvement (SDRi) measured according to \cite{vincent2006sdr} on both tasks. For RAVDESS2Mix, the results in the ``Avg.'' column indicate the SDRi of BSS averaged on both the target emotional and interference speech signals. We see that the averaged SDRi of BSS on Libri2Mix and RAVDESS2Mix are similar, which confirms the validity of our experimental setup, i.e., training on LibriMix and testing on RAVDESS2Mix.

The other columns show the SDRi of the estimated target speech. For TSE, we split the results into the cases when the emotion in the enrollment is matched or mismatched with the emotion in the target utterance.  The results of TSE for RAVDESS2Mix are in the same range as Libri2Mix when the target and enrollment are from matched emotions, and also similar to the SDRi obtained with BSS. However, we observe close to 2 dB degradation in the mismatched case. These results clearly show that emotion has a significant impact on TSE. 



To get more insights, we present the results for individual combinations of emotions in Table~\ref{tab:emo_bss_tse}. The diagonal in the TSE part of the table (highlighted with a pink background) shows the results for the matched cases. We observe that, in the matched cases, emotion slightly hurts the performance compared to the \textit{neutral-neutral} case. A similar trend is present in BSS results, where changing emotion degrades the performance by up to~1.2~dB. In the mismatched case, for TSE, the impact is much stronger. This is evident, especially in cases, where the target utterance is neutral or calm, while the enrollment contains strong emotion (e.g. calm-fearful resulting in 6.71 dB SDRi). Interestingly, we observe that the results are not symmetrical between enrollment and target emotions. For example, the SDRi fluctuates less when the enrollment is neutral or calm and the target emotion in the mixture changes than when the enrollment emotion changes and the target is neutral or calm. This suggests that in practice, it would be better to collect enrollment utterances without strong emotions or maybe to increase variation of emotions in enrollment during training.

We can conclude from this experiment that emotions affect only moderately BSS performance for all eight emotions we investigated. In contrast, some emotions greatly affect TSE performance especially when there is a mismatch between the enrollment and the target speech in the mixture. 

\subsection{Analysis of speaker identification performance}
As mentioned in Section~\ref{sec:task}, the TSE task has to solve two problems: first, identifying the target speaker in the mixture, and second, the separation itself leading to the extraction of the target signal. From the comparison of the matched and mismatched TSE cases, together with the BSS performance, we hypothesize that the issue with the emotional speech comes from the speaker identification part of the task. To support this hypothesis and to analyze this behavior further, we performed speaker verification experiments on the emotional speech.

For the speaker verification experiments, we created trials of all possible pairs of utterances in the RAVDESS dataset. We extracted embeddings $\mathbf{e}_t$ from these utterances using the auxiliary network of the TSE system $f^{(emb)}$. The embeddings are then compared using cosine distance\footnote{Note that with Euclidean distance, we obtained the same trends.}. The left part of Table~\ref{tab:sid_spkbeam} shows the equal error rate (EER) results of this experiment for different pairs of emotions in the trials. Due to the limited space, we selected four representative emotions. The performance of the speaker verification decreases steeply with the presence of emotions different from neutral. This indicates that the embeddings extracted by the TSE system are strongly sensitive to emotions, which in many cases obfuscates the speaker information. This explains the degraded performance of TSE when the enrollment contains emotion, i.e., if the speaker information in the embedding is obfuscated, the TSE network might extract the incorrect speaker. 

Some variants of TSE frameworks use external speaker embeddings trained on large corpora with the task of speaker classification \cite{wang2020voicefilter,li2019tenet}. Potentially, such embeddings might be more robust and deal better with intra-speaker variability, such as emotions. To get an idea of whether using such embeddings could solve the problem with the emotional speech, we repeated the above speaker verification experiment with x-vectors. We trained a~standard extraction network based on the~ResNet backbone~\cite{resnet}. The~network is trained on the~VoxCeleb2-dev~\cite{nagrani19} with noise and reverberation augmentations. The results are shown in the right part of Table~\ref{tab:sid_spkbeam}. The absolute speaker verification performance with x-vectors\footnote{In our notation, x-vectors refer to speaker embeddings extracted by ResNet.} is about twice better than with the embedding of the TSE system --- this is expected, as the embeddings of the TSE system were not explicitly trained for speaker verification and might not be well suited for cosine simularity metric. More interestingly, the performance of x-vectors is also significantly influenced by emotions. The degradation is smaller in the cases when the emotions in the trial are matched. In the mismatched case, the relative difference to the neutral-neutral case is close to the embeddings of the TSE system (about 4-6 times worse). This suggests that using x-vectors for TSE has the potential to slightly improve the performance on the emotional data, but most probably would not solve the problem completely.

Our experiments revealed that there is a need to increase the robustness of TSE systems to emotional speech. One option could be to learn speaker embeddings more robust to emotions by, e.g., augmenting the training data with real or synthetic emotional speech. 
An alternative option could be to exploit other clues than enrollment to identify the target speaker such as video~\cite{ephrat2018looking,afouras2018conversation,ochiai2019multimodal}, which may be less sensitive to emotions. We could exploit the visual resource of RAVDESS dataset to invesitgate if  audio-visual features could improve the robustness of TSE to emotions. 

\begin{table}[]
    \centering
    \caption{Speaker verification with TSE and x-vector embeddings. Results are in terms of EER [\%].}
    \label{tab:sid_spkbeam}
    \begin{tabular}{lc@{ }c@{ }c@{ }cc@{ }c@{ }c@{ }c}\toprule
    & \multicolumn{4}{c}{TSE embeddings} & \multicolumn{4}{c}{x-vector embeddings}\\
     \cmidrule(lr){2-5} \cmidrule(lr){6-9}
         & \textit{neutr.} & \textit{calm} & \textit{angry} & \textit{surpr.} & \textit{neutr.} & \textit{calm} & \textit{angry} & \textit{surpr.}\\\midrule
         \textit{neutral} & 4.9 & 14.2 & 31.4 & 24.1 & 2.4 & 5.9 & 11.1 & 9.8\\
         \textit{calm} & 14.2 & 9.6 & 30.3 & 25.2 & 5.9 & 7.3 & 12.9 & 11.4\\
         \textit{angry} & 31.4 & 30.3 & 30.4 & 31.6 & 11.1 & 12.9 & 9.0 & 12.2\\
         \textit{surprised} & 24.1 & 25.2 & 31.6 & 19.3 & 9.8 & 11.4 & 12.2 & 7.3 \\\bottomrule
    \end{tabular}
\end{table}

\section{Conclusion}
\label{sec:conclusion}
In this work, we investigated the influence of emotions on TSE and BSS. For this purpose, we created a new emotional speech mixture evaluation dataset called RAVDESS2Mix, which enabled us to perform controlled experiments to analyze the impact of eight emotions on the performance of BSS and TSE. 
Our experiments revealed that emotions had a relatively small impact on BSS but a larger impact on TSE, especially in the case of mismatched emotions between the enrollment and the target speech in the mixture. 
We hypothesized that this performance degradation is mainly due to the difficulty in identifying the target speaker, which we confirmed with speaker verification experiments. 
We hope that the results of this study will foster more research on solutions to increase the robustness of TSE systems to internal speaker variability such as emotions.


\section{Acknowledgement}
\label{sec:acknowledgement}
The work was partly supported by European Union’s Horizon 2020 project No. 870930 --- WELCOME, and by Czech Ministry of Education, Youth and Sports from project no. LTAIN19087 "Multi-linguality in speech technologies". Computing on IT4I supercomputer was supported by the Czech Ministry of Education, Youth and Sports from the Large Infrastructures for Research, Experimental Development and Innovations project "e-Infrastructure CZ --- LM2018140".


\bibliographystyle{IEEEbib}
\bibliography{refs}

\end{document}